%% file: ANA-EXOT-2018-09-PAPER.tex
\documentclass[cernpreprint, atlasdraft=false, texlive=2023, USenglish, texmf, orcidlogo]{atlasdoc}
 
\usepackage{atlaspackage}
\usepackage{atlasbiblatex}
\usepackage{amsmath}
 
\usepackage{atlasphysics}
 
\graphicspath{{logos/}{figures/}}
 
\usepackage{ANA-EXOT-2018-09-PAPER-defs}
 
\newcommand{\Zprime}{\ensuremath{Z^{\prime}}\xspace}
\newcommand{\Wprime}{\ensuremath{W^{\prime}}\xspace}
\newcommand{\mjj}{$m_{\text{jj}}$\xspace}
\DeclarePairedDelimiterX{\infdivx}[2]{}{}{
#1\;\delimsize\|\;#2
}

\makeatletter
\DeclareOldFontCommand{\rm}{\normalfont\rmfamily}{\mathrm}
\DeclareOldFontCommand{\sf}{\normalfont\sffamily}{\mathsf}
\DeclareOldFontCommand{\tt}{\normalfont\ttfamily}{\mathtt}
\DeclareOldFontCommand{\bf}{\normalfont\bfseries}{\mathbf}
\DeclareOldFontCommand{\it}{\normalfont\itshape}{\mathit}
\DeclareOldFontCommand{\sl}{\normalfont\slshape}{\@nomath\sl}
\DeclareOldFontCommand{\sc}{\normalfont\scshape}{\@nomath\sc}
\makeatother

\AtlasTitle{Search for heavy particles in the $b$-tagged dijet mass distribution with additional $b$-tagged jets in proton-proton collisions at $\sqrt{s} = 13$ TeV with the ATLAS experiment}

\AtlasAbstract{ A search optimized for new heavy particles decaying to two
$b$-quarks and produced in association with additional $b$-quarks is reported.
The sensitivity is improved by $b$-tagging at least one
lower-$p_{\rm{T}}$ jet in
addition to the two highest-$p_{\rm{T}}$ jets. The data used in this search correspond
to an integrated luminosity of 103~$\text{fb}^{-1}$ collected with a dedicated
trijet trigger during the 2017 and 2018 $\sqrt{s} = 13$ TeV proton--proton
collision runs with the ATLAS detector at the LHC. The search looks for
resonant peaks in the $b$-tagged dijet invariant mass spectrum over a smoothly
falling background.  The background is estimated with an innovative data-driven
method based on orthonormal functions. The observed $b$-tagged dijet invariant
mass spectrum is compatible with the background-only hypothesis. Upper limits
at 95\% confidence level on a heavy vector-boson production cross section times
branching ratio to a pair of $b$-quarks are derived.}
 
\author{The ATLAS Collaboration}

\PreprintIdNumber{CERN-EP-2021-119}
\AtlasJournalRef{Phys. Rev. D 105 (2022) 012001}
\AtlasDOI{10.1103/PhysRevD.105.012001}

\hypersetup{pdftitle={ATLAS document},pdfauthor={The ATLAS Collaboration}}
 
\begin{document}
 
\maketitle

% The next lines are included from the .//chapters/Introduction.tex input file
\section{Introduction}
\label{sec::intro}
 
\begin{figure}[H]
\centering
\includegraphics[width=0.6\textwidth]{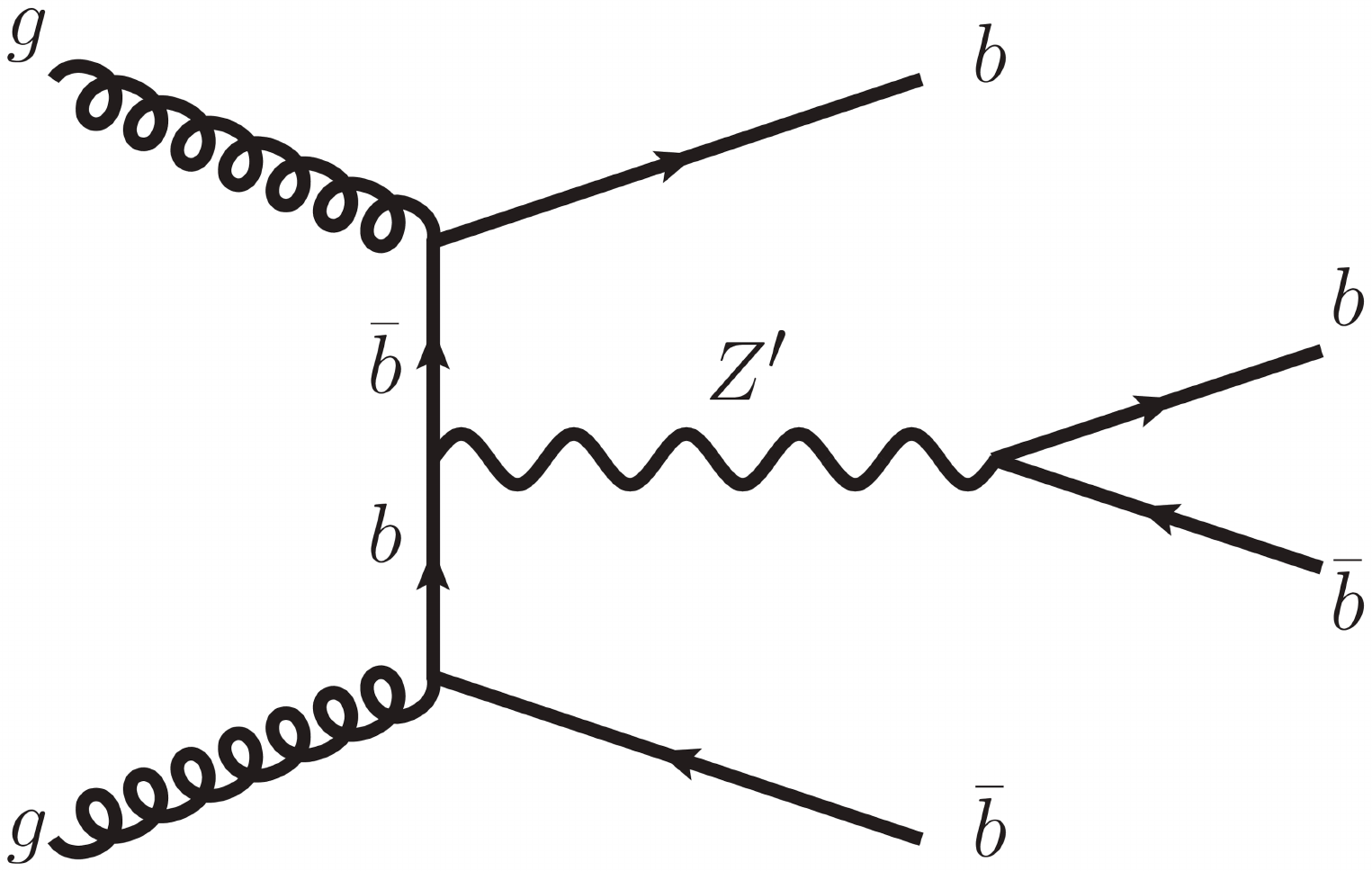}
\caption{Representative Feynman diagram for the resonant production of a \Zprime\ boson in association with additional $b$-quarks and its decay into a pair of $b$-quarks.}
\label{fig:ediag}
\end{figure}

New particles beyond the Standard Model with preferential couplings to
third-generation quarks are predicted by many
models~\cite{Kim:2016csm,Jacques:2016dqz,HILL1995483,Balazs:1998nt,PhysRevLett.83.3370,PhysRevD.91.015017}
and have been extensively searched for by experiments at the Tevatron and the LHC.
Searches for heavy particles in the dijet invariant mass spectrum of ~``$b$-tagged'' jets,
identified as containing $b$-hadrons, have probed this
scenario~\cite{Sirunyan:2018pas,CMSDiB,DiJet2020Paper,DiJetISR2016Paper, bHbb,Sirunyan:2018taj}.
Recent results from the LHCb and Belle experiments suggest deviations in lepton-flavor
universality (LFU) from Standard Model (SM) expectations
~\cite{Aaij:2017uff,Hirose:2017dxl,Aaij:2017vbb,Aaij:2018jhg,Aaij:2021vac}, referred to as
LFU anomalies. The phenomenological work in
Refs.~\cite{Greljo:2015mma,Faroughy:2016osc} predicts new heavy vector bosons, a
\Zprime\ boson and \Wprime boson, mainly coupled to third-generation
leptons and quarks. This model not only offers an explanation for the
LFU anomalies, but also predicts \Zprime\ boson production in association with
third-generation quarks via gluon splitting in the LHC experiments, due to the vanishingly small
$b$-quark and top-quark parton distribution functions (PDFs) for protons at the
\TeV\ scale. The dominant Feynman diagram for this $b$-quark associated production
is shown in Figure~\ref{fig:ediag}.
 
\label{sec:intro}
 
The additional $b$-flavor jets, not coming from the heavy-particle decay, can
be used to reduce the multijet background contribution since the additional
jets in multijet events most likely do not contain $b$-hadrons. Most previous
searches for heavy particles in dijet final states did not require additional
$b$-tagged jets beyond the leading two jets~\cite{CMSDiB,Sirunyan:2018pas,
DiJet2020Paper,DiJetISR2016Paper}, i.e. the two jets with the highest
transverse momentum (\pT). This article presents a search for heavy particles
in final states where the two leading jets are $b$-tagged and either the third
or fourth jet is also $b$-tagged.  The $b$-tagging criterion for the third and
fourth jets increases the sensitivity by 20\%--50\% at a mass scale of
1.3~\TeV\ to 3~\TeV. This is the first search probing the mass region
up to 3.6~\TeV\ in this final state. Previous searches for a heavy Higgs
boson~\cite{bHbb,Sirunyan:2018taj} in a similar final state were optimized for
masses up to 1.4~\TeV. An advanced $b$-tagging algorithm with
better performance at high \pt~\cite{Aad:2019aic} is applied in this search
compared to the one used in Ref~\cite{bHbb}, enhancing the sensitivity in the
high mass region. The model proposed in
Refs.~\cite{Greljo:2015mma,Faroughy:2016osc} is compared with data for the
first time.

% End of text imported from the .//chapters/Introduction.tex input file

\section{ATLAS detector}
\label{sec:detector}
 
The ATLAS detector~\cite{Collaboration_2008} at the LHC covers nearly the
entire solid angle around the collision point.\footnote{ATLAS uses a
right-handed coordinate system with its origin at the nominal interaction point
(IP) in the center of the detector and the $z$-axis along the beam pipe. The
$x$-axis points from the IP to the center of the LHC ring, and the $y$-axis
points upward. Cylindrical coordinates ($r,\phi$) are used in the transverse
plane, $\phi$ being the azimuthal angle around the $z$-axis. The pseudorapidity
is defined in terms of the polar angle $\theta$ as $\eta = -\ln[\tan(\theta/2)]$.
The rapidity is defined as $y=(1/2)\ln[(E+p_{z})/(E-p_{z})]$, where $E$ is the
energy and $p_{z}$ is the momentum in the $z$ direction.
Transverse energy is defined as $\ET = E\sin\theta$.} It consists of an inner tracking detector surrounded by a thin
superconducting solenoid, electromagnetic and hadronic calorimeters, and a muon
spectrometer. The inner-detector system is immersed in a 2 T axial magnetic
field and provides charged-particle tracking in the range $|\eta| < 2.5$. The
high-granularity silicon pixel detector covers the vertex region and typically
provides four measurements per track, the first hit normally being in the
insertable B-layer installed before Run~2~\cite{Capeans:1291633,Abbott:2018ikt}. It is followed by the silicon
microstrip tracker, which usually provides eight measurements per track. These
silicon detectors are complemented by the transition radiation tracker, which
enables radially extended track reconstruction up to $|\eta| = 2.0$ and
contributes to electron identification. The calorimeter system covers the
pseudorapidity range $|\eta| < 4.9$. Within the region $|\eta| < 3.2$,
electromagnetic calorimetry is provided by barrel and endcap high-granularity
lead/liquid-argon (LAr) calorimeters, with an additional thin LAr presampler
covering $|\eta| < 1.8$ to correct for energy loss in material upstream of the
calorimeters. Hadronic calorimetry is provided by the steel/scintillator-tile
calorimeter, segmented into three barrel structures within $|\eta| < 1.7$, and
two copper/LAr hadronic endcap calorimeters. The solid angle coverage is
completed with forward copper/LAr and tungsten/LAr calorimeter modules
optimized for electromagnetic and hadronic measurements, respectively. The
outermost layers of ATLAS consist of a muon spectrometer within $|\eta| < 2.7$,
incorporating three large superconducting toroidal magnet systems.  Interesting
events are selected by the first-level trigger system implemented in custom
hardware, followed by selections made by algorithms implemented in software in
the high-level trigger computer farm~\cite{Aaboud:2016leb,Aad:2020wji}. The
first-level trigger accepts events from the 40~MHz bunch crossings at a rate
below 100~kHz, which the high-level trigger reduces in order to record events
to disk at a rate of about 1~kHz. An extensive software
suite~\cite{ATL-SOFT-PUB-2021-001} is used for real and simulated data
reconstruction and analysis, for operation and in the trigger and data
acquisition systems of the experiment.

% The next lines are included from the .//chapters/Samples.tex input file
\section{Simulated event samples}
\label{sec:samples}
 
The theory model used in this search considers a \Zprime\ boson exclusively
coupled to the third generation of the left-handed SM
fermions~\cite{Greljo:2015mma,Faroughy:2016osc}. Events with a \Zprime\ boson
produced in association with up to two additional partons in the final state,
with the coupling to the quark sector ($g_{b}$) set to unity and the coupling
to the lepton sector ($g_{\tau}$) set to zero, were generated using
\MGMCatNLOV{2.2.4}~\cite{MadGraph5}, considering all diagrams at the
matrix-element level of the type $pp\rightarrow Z', pp\rightarrow Z'j$ and
$pp\rightarrow Z'jj$, and the parton showering was done in
\PYTHIAV{8.212}~\cite{Pythia8}. Both $Z'\rightarrow t\overline{t}$ and
$Z'\rightarrow b\overline{b}$ are possible with $g_{b}$ set to unity, but only
the latter was included in the generation. The five-flavor scheme leading-order
(LO) NNPDF3.0 PDF~\cite{Ball:2012cx} and a set of tuned parameters called the
A14 tune~\cite{ATL-PHYS-PUB-2014-021} were applied in the generation. CKKW-L
$k_{t}$-merging ~\cite{Lonnblad:2001iq} was used to match the multileg
parton-level events with \PYTHIAV{8}. The intrinsic width of the \Zprime\ boson
is roughly 4\% of the \Zprime\ boson mass. This model is referred to as the
lepton-universality-violating (LUV) \Zprime in this article.
 
Multijet events were generated with \PYTHIAV{8.186}, using the LO NNPDF2.3
PDF~\cite{Martin:2009iq} and the A14 tune. The
background is estimated in a fully data-driven way and the simulated multijet
event samples are only used to optimize the event selections.
 
The simulated multijet events were passed through a full ATLAS detector
simulation~\cite{Aad:2010ah} using $\GEANT4$~\cite{AGOSTINELLI2003250}. The
signal samples were passed through a fast simulation where the response in the
calorimeters is provided by a parametrization~\cite{ATLAS:1300517} while
$\GEANT4$ is used elsewhere. The decay of $b$- and $c$-hadrons was performed
using the \EVTGEN ~1.2.0 decay package~\cite{LANGE2001152}. Additional
proton--proton interactions (pileup) from the same and neighboring bunch
crossings were taken into account by generating a number of inelastic $pp$
interactions with \PYTHIAV{8.186} using the LO NNPDF2.3 PDF set and the A3
tune~\cite{ATL-PHYS-PUB-2016-017}.  These events were then overlaid with the
hard-scattering events. All simulated events are weighted so that the
distributions of the average number of collisions per bunch crossing match
those in data.

% End of text imported from the .//chapters/Samples.tex input file

% The next lines are included from the .//chapters/DataAndSelections.tex input file
\section{Data sample and event selection}
\label{sec:dataAndSel}
 
The data used for this analysis were collected with the ATLAS detector from
proton--proton ($pp$) collisions at the LHC with a center-of-mass energy of
$\sqrt{s}=13$ \TeV\ in 2017 and 2018, requiring that all detector systems were
functional and recording high-quality data~\cite{DQ}. The dataset corresponds
to an integrated luminosity of 103~\ifb~\cite{ATLAS-CONF-2019-021}, using the
LUCID-2 detector~\cite{LUCID2} for the primary luminosity measurements. Data
were collected using a newly developed trijet trigger with asymmetric
thresholds, first deployed in 2017. This trigger required two jets to have
transverse energy \ET\ greater than 250~\GeV\ and a third jet to have \ET\
greater than 120~\GeV. All three jets were required to have $|\eta| < 3.2$,
with intermediate preselections at lower \pt\ requiring $|\eta| < 2.4$.  This
trigger allows this search to reach an invariant mass (\mjj) of the two
highest-\pT\ jets about 100~\GeV\ lower than a single-jet trigger would allow.
The trigger efficiency exceeds 99.95\% after applying the kinematic selections
on jets described below.
 
Collision vertices are reconstructed from at least two tracks with $\pT\ >0.5$
\GeV. The one with the highest $\sum \pTsq$\ of the associated tracks is
considered to be the primary vertex. Jets are reconstructed using the
anti-$k_{t}$ algorithm~\cite{Cacciari:2008gn, Cacciari:2008gp,Cacciari:2011ma} with a radius
parameter of $R=0.4$ from noise-suppressed topological energy
depositions~\cite{Aad:2016upy}. A calibration sequence is applied to jet
energies and directions as described in
Refs.~\cite{atlascollaboration2020jet}. Jets with $\pT > 25$~\GeV\ are
removed if they are compatible with noise bursts, beam-induced background or
cosmic rays with the ``loose'' criteria defined in Ref.~\cite{jetclean}.
 
A deep-learning neural network, DL1r, is applied to identify jets containing
$b$-hadrons~\cite{Aad:2019aic}. This algorithm utilizes the distinctive
characteristics of $b$-hadron decays such as the large impact parameters of
tracks and the displaced vertices reconstructed in the inner detector.
Therefore it is only applied to jets with $|\eta| < 2.5$. In addition, it
includes the output from a recurrent neural network (RNNIP)~\cite{RNNNote},
which exploits the correlations between tracks originating from the same
$b$-hadron. Operating points (OP) are defined by a single cut-value on the
discriminant output distribution and are chosen to yield a certain \btag\
efficiency in a simulated inclusive $t\overline{t}$ event sample. The 77\%
efficiency \btag\ OP is selected as it gives the best overall signal
sensitivity across the \mjj\ range under consideration. In simulated multijet
events, the efficiency drops from 77\% on average for $b$-jets with $\pT <
250$~\GeV\ to 10\% for a \pT\ of 2 \TeV. The corresponding mis-tag rate for
charm-flavor (light-flavor) jets changes from 20\% ($<1\%$) to 2\% ($<1\%$).
 
The analysis considers events with at least four jets. The leading two jets
must have $\pT > 250$~\GeV\ and the third leading jet must have $\pT >
120$~\GeV. The leading three jets must be within $|\eta| < 2.4$ to match the
preselections applied in the trigger chain. The fourth leading jet must have
$\pT > 25$~\GeV\ and be within $|\eta| < 4.5$. To suppress low-\pT\ jets from
underlying events, a multivariate classification algorithm based on tracking
information, the jet vertex tagger (JVT), is applied and jets within $|\eta| <
2.5$ with $\pT < 60$ ~\GeV\ must satisfy the ``medium''
OP~\cite{ATLAS:2015ull}. Jet pairs from the decay of heavy particles are more
central than those from multijet events, thus the rapidity separation of two
leading jets, defined as $y^{\ast} = (y_{1}-y_{2})/2$, is a powerful
discriminant. The optimal choice is found to be $|y^{\ast}| < 0.8$. The \mjj\
is required to be greater than 730~\GeV\, above which the spectrum starts to
decrease continually. An event is preselected if it passes all the above requirements.
 
The signal region (SR) selection requires both the leading two jets and either
the third or fourth jet to satisfy the 77\% $b$-tagging OP. The full selection
is summarized in Table~\ref{tab:selections}. The acceptance of the benchmark
signal is 1.9\%, 2.4\% and 1.1\% for an LUV \Zprime mass of 1.3, 1.6 and
2.5~\TeV\ respectively, rising then falling due to increasing kinematic
acceptance and decreasing $b$-tagging efficiency with jet \pT. The
corresponding \mjj spectra after applying the full selection are shown in
Figure~\ref{fig:SignalShape}. The \mjj distributions are wide due to the
significant low \Zprime\ mass tails introduced by off-shell production, which occurs due
to the steeply falling PDF of the two colliding partons at large values of
Bjorken $x$. This feature is commonly referred to as a~``parton-luminosity
tail'', and its size increases with the resonance mass. This effect is more
significant for LUV \Zprime as the $b$-quark PDF has a larger proportion at low
Bjorken $x$ values than those of light quarks, $c$-quarks and gluons.
 
\begin{table}[htbp]
\begin{center}
\caption{Signal region event selection criteria. The superscripts refer to the ranking given by jet \pT.}

% The next lines are included from the .//tables/Selections.tex input file
\makebox[0pt]{
\renewcommand{\arraystretch}{1.2}
\begin{tabular}{|c|c|}
\hline
\multicolumn{2}{|c|}{Preselection}\\
\hline
\multirow{3}{*}{Jet \pT} & $\pT^{1,2} > 250$ \GeV \\
& $\pT^{3} > 120$ \GeV\\
& $\pT^{4} > 25$ \GeV\\
\hline
\multirow{2}{*}{Jet $|\eta|$} & $|\eta^{1,2,3}| < 2.4$\\
& $|\eta^{4}| < 4.5$ \\
\hline
JVT &``medium'' OP \\
\hline
$|y^{\ast}|$ & $< 0.8$\\
\hline
\multicolumn{2}{|c|}{$b$-tagging Selection}\\
\hline
\multirow{2}{*}{DL1r} & Leading two jets: 77\% OP\\
& Third or fourth jet: 77\% OP\\
\hline
\mjj & $> 730$ \GeV\\
\hline
\end{tabular}
}

% End of text imported from the .//tables/Selections.tex input file

\label{tab:selections}
\end{center}
\end{table}
 
\begin{figure}[!thb]
\centering
\includegraphics[scale=0.5]{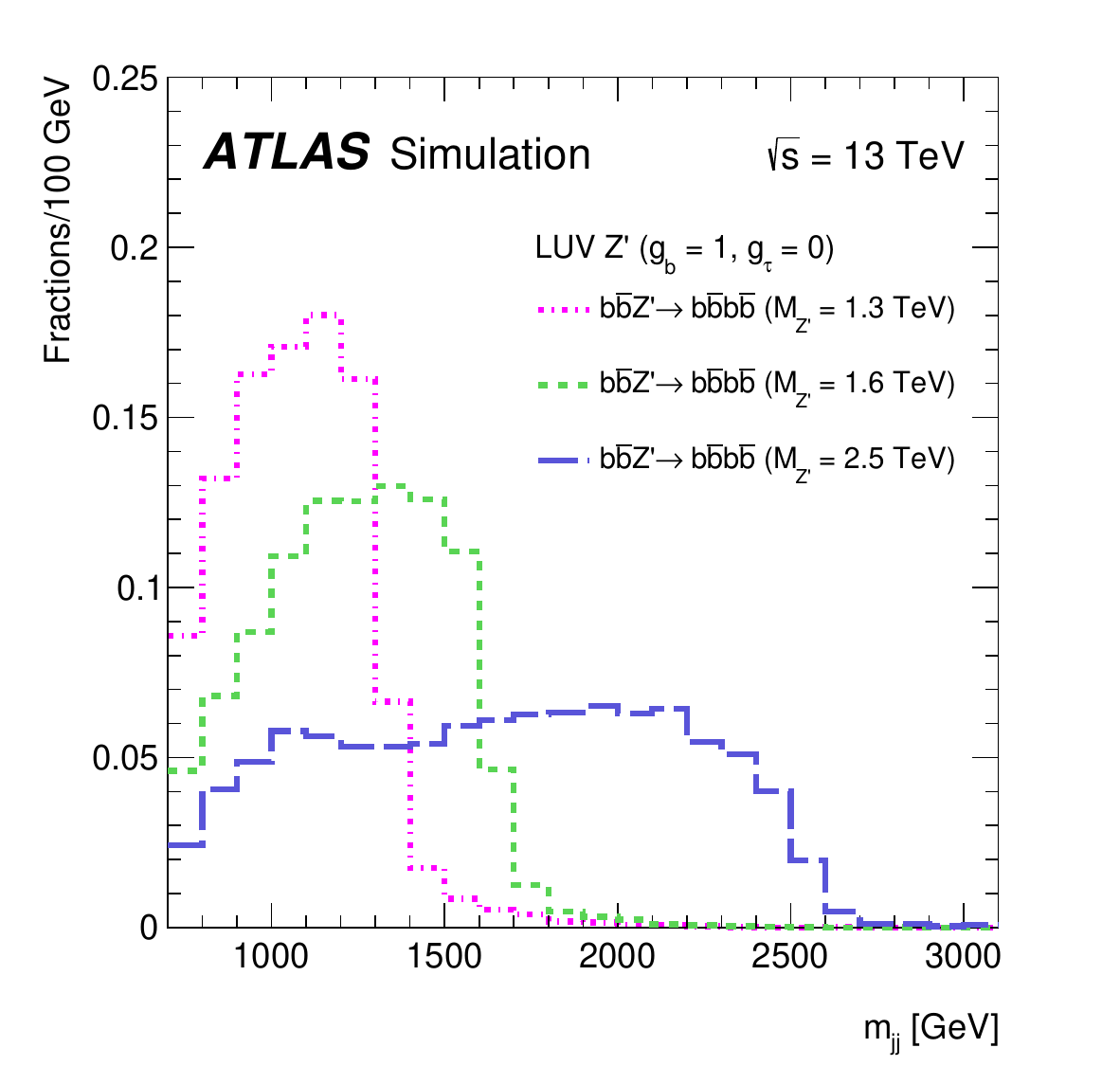}
\caption{\mjj distributions of three benchmark LUV \Zprime mass points, 1.3~\TeV, 1.6~\TeV\ and 2.5~\TeV\, after applying the full signal region selections, normalized to unity.}
\label{fig:SignalShape}
\end{figure}

% End of text imported from the .//chapters/DataAndSelections.tex input file

% The next lines are included from the .//chapters/Backgrounds.tex input file
\section{Background estimation}
\label{sec:bkg}
 
The main background in this search originates from multijet events. A novel
data-driven method is used to estimate the background contribution by
performing a fit to the \mjj\ spectrum. The shape of this spectrum is
impacted by the asymmetric thresholds of the trijet trigger used to collect
the data. The empirical functional forms used in previous heavy-particle
searches~\cite{DiJet2020Paper} are not able to model this effect adequately.
The background is estimated with the ``functional decomposition'' (FD)
method~\cite{fd_paper,ATL-PHYS-PUB-2020-028}, which
uses a truncated series. First it performs a power-law transformation on \mjj,
as given by
 
\begin{equation*}
z \equiv \left(\frac{m_{\text{jj}} - m_{\text{jj}}^{0}}{\lambda}\right)^{\alpha},
\label{equ:exp_trans}
\end{equation*}
 
where $m_{\text{jj}}^{0}$ is the start of the spectrum (730~\GeV), $\lambda$ is a
positive scale factor and $\alpha$ is a positive exponent. The spectrum can then be
modeled by
 
\begin{equation*}
\Omega(z) = \sum\limits_{n=1}^{N}c_{n}E_{n}(z),
\label{equ:trunc_series}
\end{equation*}
 
where $E_{n}(z)$ is the $n$-th member of an orthonormal basis constructed from
exponential functions, ${c_{n}}$ is the corresponding $n$-th coefficient and
$N$ is the total number of terms used. Equation~(\ref{func:moments}) shows the
first three orthonormal functions.
 
\begin{equation}
\begin{aligned}
E_{1}(z) &= \sqrt{2}e^{-z},\\
E_{2}(z) &= 6e^{-2z} - 4e^{-z},\\
E_{3}(z) &= 10\sqrt{6}e^{-3z} - 12\sqrt{6}e^{-2z} + 3\sqrt{6}e^{-z}.
\end{aligned}
\label{func:moments}
\end{equation}
 
Each pair of ($\lambda$, $\alpha$) defines a hyperspace of functions. For a
given pair of ($\lambda$, $\alpha$), the \mjj\ spectrum can be
modelled precisely as $N\to\infty$. Since a falling \mjj\ spectrum can be
described adequately using lower moments, while a localized structure in \mjj\
needs higher moments to model it well, truncating the series at a certain $N$
allows the background to be modeled mostly independently of the presence of
signal~\cite{fd_paper}. There exists an optimal set of ($\lambda$, $\alpha$,
$N$) that gives the most succinct representation of the data. It is identified
by using two steps: dataset decomposition and parameter optimization. In the
dataset decomposition step, the \mjj\ spectrum is converted to an
unbinned dataset and the FD method creates a complete representation of this
dataset in an initial hyperspace defined by ($\lambda_0$, $\alpha_0$),
considering 4096 moments as an approximation of $N\to\infty$. Values of $c_{n}$
and the covariance matrix are calculated. The parameter optmization step
performs a grid scan on the $\lambda$--$\alpha$ plane followed by a
gradient-descent minimization starting from the best point obtained in the grid
scan. The grid scan covers a large space, with $\alpha$ ranging from 0.6 to 1.3
and $\lambda$ ranging from 50 to 750. Each axis has 140 evenly distributed
steps and there are 19600 points scanned in total. The starting point is (0.6,
50). The covariance matrix and $c_{n}$ in another hyperspace can be obtained by
a matrix transformation between the initial hyperspace and the new
one~\cite{fd_paper}, thus the dataset decomposition is only carried out once at
the starting point.  Only the first 128 moments are considered in the parameter
optimization step as it is already well above the number of moments needed for
the spectrum under consideration. Every combination of ($\lambda$, $\alpha$) and $N$
is considered and the quantity to minimize is~\cite{fd_paper}:
 
\begin{equation}
\mathcal{L} = D_{\mathrm{KL}}(\tilde{\mathbf{f}}\lVert\tilde{\mathbf{c}}) +  D_{\mathrm{KL}}(\tilde{\mathbf{c}}\lVert\tilde{\mathbf{p}}),
\label{equ:likelihood}
\end{equation}

which contains two Kullback--Leibler divergences ($D_{\mathrm{KL}}$)~\cite{KL}
and measures how a probability distribution differs from a reference
probability distribution. The first term is the divergence of the estimate
($\tilde{\mathbf{c}}$) with respect to data ($\tilde{\mathbf{f}}$). A better
description of the data results in a smaller divergence. The second term is the
divergence of some prior background assumption ($\tilde{\mathbf{p}}$) with
respect to the estimate ($\tilde{\mathbf{c}}$). It is a penalty term as
$\tilde{\mathbf{p}}$ is constructed such that the divergence increases as the
value of $N$ increases~\cite{fd_paper}. As a consequence, the minimization
prefers a smaller number of terms and the series is truncated. The truncated
series, given by the optimized ($\lambda$, $\alpha$), $N$ and the corresponding
coefficients ($c_{n}$), determines the background estimate. Even though the
penalty term in Eq.~(\ref{equ:likelihood}) prevents overfitting the data, the
upper limit on $N$ is fixed using background-only pseudodata samples introduced
below. A lower limit on $N$ is also identified using the same pseudodata
samples to ensure the background estimate is adequate. In the application to
data, the parameter optimization step only considers values of $N$ within this
fixed range while other parameters are not constrained by background-only
pseudodata samples. The background estimate is converted to a histogram as the
input to the statistical analysis, where the binning is determined by the
\mjj\ resolution measured in Ref.~\cite{DiJet2020Paper}.
 
Background-only pseudodata samples are generated by scaling the events passing
the preselections listed in Table~\ref{tab:selections} with the expected
event-level $b$-tagging selection efficiency of multijet events in data. The
event-level $b$-tagging selection efficiency is defined as the probability, as
a function of the \mjj, of a preselected event to enter the signal
region. The measured efficiency points are fitted by a third-order polynomial
to smooth any features introduced by local fluctuations or by the presence of
signal events. It is verified that a third-order polynomial is insensitive to
possible signal contamination by artificially introducing different signal
width and mass combinations. The event-level $b$-tagging selection efficiency
for the target signal region is calculated using data to be 0.14\% at 730~\GeV\
and decreases to 0.01\% at 3~\TeV. Pseudodata samples are generated by varying
the fitted efficiency curve according to the uncertainties associated with the
fit.
 
A background-enhanced control region with negligible signal contamination,
where both the third and fourth jets are required to fail the $b$-tagging
requirements, is constructed to validate this procedure for generating
pseudodata samples. In this control region, the corresponding event-level
$b$-tagging selection efficiency is the probability that an event fails to
satisfy the $b$-tagging criteria for the third and fourth jets but has both the
leading two jets $b$-tagged. The pseudodata samples generated by this procedure
provide a good description of the target spectrum in this control region.
 
The background-only pseudodata samples approximating the signal region are used
to check whether a given FD configuration models the background adequately by
performing two sets of tests. A $\chi^{2}$ test is applied to check whether the
estimate agrees well with the background. More than 90\% of the
pseudoexperiments are required to have a $\chi^{2}$ $p$-value greater than
0.01. In addition to the $\chi^{2}$ test, the \texttt{BumpHunter}
algorithm~\cite{Aaltonen:2008vt,Choudalakis:2011qn} described in
Section~\ref{sec:results} is used to check whether the background estimate
introduces strong biases in the statistical analysis. More than 90\% of the
pseudoexperiments must give a \texttt{BumpHunter} $p$-value greater than 0.01
when the background-only pseudodata samples are considered.
 
The lower and upper limits on $N$ are determined by the above tests, starting
without any constraints on $N$ and gradually shrinking the allowed range. First
the lower limit is optimized by increasing it until the above criteria are met.
Once the lower limit is determined, the upper limit is decreased until the
above criteria are no longer satisfied. The tests show that requiring $N$ to be
2 or 3 is sufficient to obtain a good estimate of the background. Therefore, the
minimization only considers these two options in the signal injection tests
described below and in the final application to data.
 
Another set of tests is performed with signal events injected into the
background-only pseudodata samples to evaluate the sensitivity to signal. The
sensitivity to a given signal model is considered to be optimal if more than
90\% of pseudoexperiments return a \texttt{BumpHunter} $p$-value less than 0.01
with a significant number of signal events injected. The number of events
injected satisfies $N_{\mathcal{S}}/\sqrt{N_{\mathcal{B}}}$ = 5, in which
$N_{\mathcal{S}}$ and $N_{\mathcal{B}}$ are the numbers of signal and
background events, respectively, within a \mjj\ window corresponding to twice
the signal width on each side of the signal mass point. A large set of Gaussian-like
signal hypotheses across the entire \mjj\ region, with various widths, are
tested; and they show that this search has optimal sensitivity above 1.3~\TeV.
The low \mjj\ region is sculpted more significantly by the trigger and kinematic
selections. The truncated series determined by FD is too flexible and absorbs
more of the signal in this region, resulting in lower sensitivities. Therefore,
the signal search is conducted only in the region \mjj\ $> 1.3$~\TeV\ despite
the whole region, \mjj\ $> 730$~\GeV, being considered in the FD. The lower
\mjj\ region is used as a sideband.

% End of text imported from the .//chapters/Backgrounds.tex input file

% The next lines are included from the .//chapters/Systematics.tex input file
\section{Systematic uncertainties}
\label{sec:sys}
 
The main systematic uncertainties in the simulated signal samples consist of
those associated with the modeling of the jet energy scale (JES), the jet
energy resolution (JER) and the $b$-tagging efficiency. JES and JER variations
are applied to the signals to obtain varied signal templates. They are
estimated using jets in $\sqrt{s} = 13$~\TeV\ data and simulation in various
methods~\cite{atlascollaboration2020jet}. The impact of these uncertainties on
the expected number of signal events across the whole \mjj\ range is 24\%-30\%
(35\%-40\%) from JES (JER) variations. The systematic uncertainty of the
$b$-tagging efficiency is measured using data enriched in $t\overline{t}$
events for jet $\pT < 400$~\GeV\ and extrapolated to higher \pT\ regions using
a method similar to the one described in Ref.~\cite{ATL-PHYS-PUB-2021-003}.
This uncertainty ranges from 12\% to 20\% depending on the reconstructed signal
mass. A luminosity uncertainty of 2.4\%~\cite{ATLAS-CONF-2019-021} is applied
to the normalization of the signal samples.
 
The background modeling uncertainties come from both the statistical
uncertainty in the fit parameters, referred to as the fit parameter
uncertainty, and the biases introduced by the fit model itself. The fit
parameter uncertainty is evaluated by propagating the covariance matrix
returned by the fit to each \mjj\ bin, giving an uncertainty ranging from 2\% to
5\% relative to the background estimate. The biases from the fit model are
evaluated with pseudoexperiments using signal-injected pseudodata samples,
quantifying how much the background estimate is biased by injected signal
events in the spectrum that are not able to create significant excesses. The
number of signal events injected corresponds to the mean of the expected limits
obtained for each signal hypothesis when background-only pseudodata samples are
considered. The background estimate is increased by 5\% to 40\% at the injected
signal mass point compared to the nominal case. This is referred to as the fit
bias uncertainty.
 
The fit model may not describe the background perfectly, so that signal-like
features can arise in data relative to the fit even if there are no signal
events in data, introducing spurious signals. A spurious-signal uncertainty is
included to address this. It is evaluated using pseudoexperiments, for each of
which a fixed background template, obtained from an FD fit to the background-only pseudodata sample,
and the signal templates have their normalizations fitted to the
background-only \mjj\ spectrum in the manner described in Section~\ref{sec:results},
and the obtained number of signal events is taken as the spurious signal. It is
measured for each signal mass point and a smooth envelope is constructed following the
method described in Ref.~\cite{ATL-PHYS-PUB-2020-028}. The size of this
uncertainty relative to the background estimate increases from 0.5\% at 1.3
~\TeV\ to 14\% at 3.0~\TeV.

% End of text imported from the .//chapters/Systematics.tex input file

% The next lines are included from the .//chapters/Interpretation.tex input file
\section{Results and interpretation}
\label{sec:results}
 
The \texttt{BumpHunter} algorithm is
adopted to quantify the statistical significance of any localized excess in the
binned \mjj distribution. It calculates the significance of any
excess found in continuous intervals in all possible locations. The search
window has a width ranging from two bins up to half of the full \mjj\ distribution. In each window it evaluates the significance of the discrepancy
between the data and the background yield. The most significant interval is the set
of bins with the smallest probability of arising from background fluctuations.
A $p$-value is reported by \texttt{BumpHunter}, corresponding to the
probability of random fluctuations in the background-only hypothesis creating
an excess at least as significant as the one observed anywhere in the spectrum.
This is obtained by performing a series of pseudoexperiments constructed from
the background estimate so that the look-elsewhere effect~\cite{Gross_2010} is taken into
account.
 
Three moments are used ($N=3$) after applying the optimization procedure
described in Section~\ref{sec:bkg}. The corresponding fit describes the data
very well across the entire \mjj\ region, obtaining $\chi^{2}$/nDoF = 0.72 and a
$\chi^{2}$ $p$-value of 0.89~\cite{1900wow}. \texttt{BumpHunter} reports a $p$-value of 0.55
after scanning the \mjj\ range starting at 1.3~\TeV\, with the most significant interval found to be $[1.92,2.11]$~\TeV, as shown in
Figure~\ref{fig:fit_plot}.
 
\begin{figure}[!thb]
\centering
\includegraphics[scale=0.5]{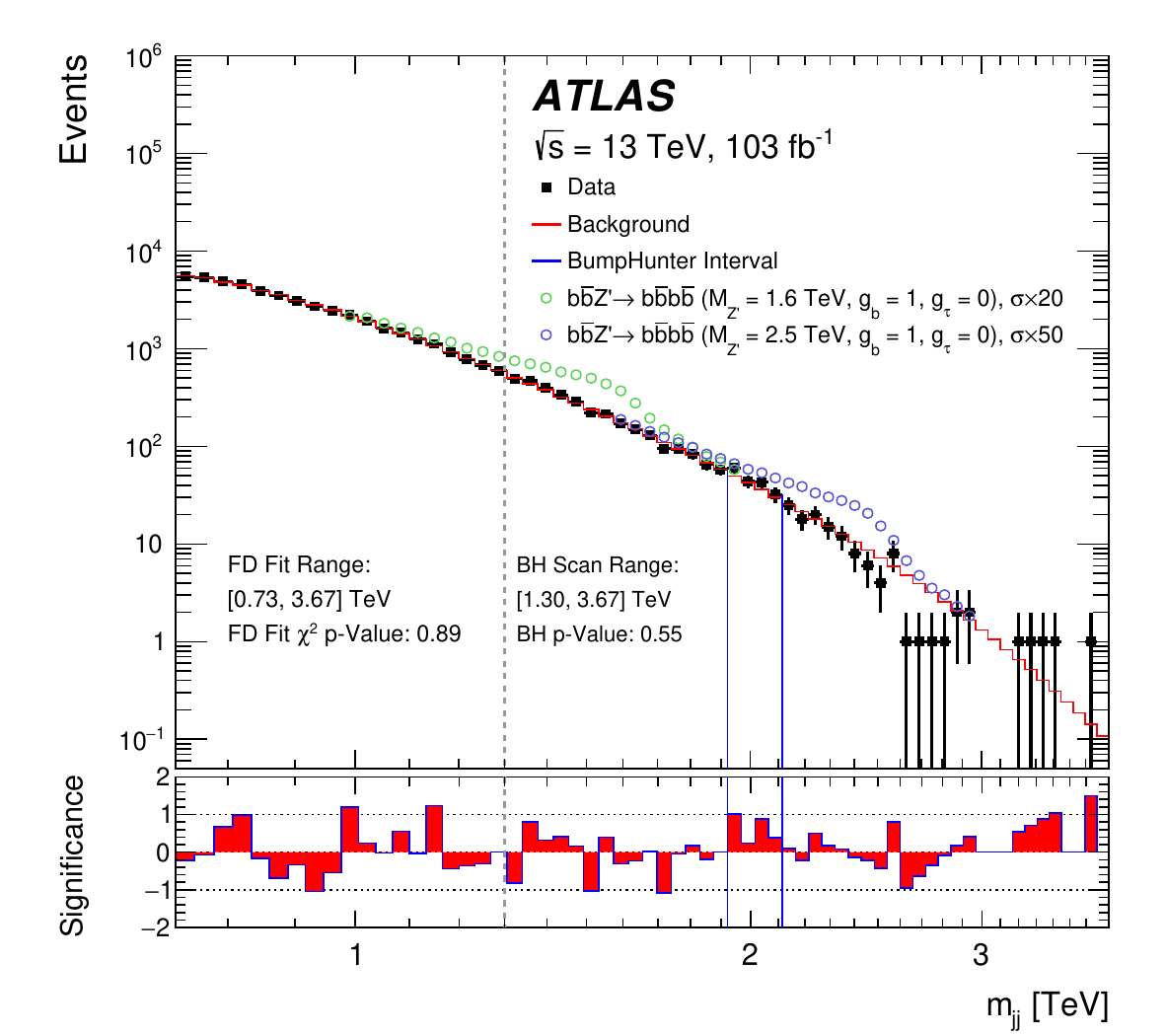}
\caption{Background estimate from the FD method with $N=3$ and data in the SR. The solid squares show the \mjj\ spectrum in data and the solid curve is the background estimated by the FD method. The \texttt{BumpHunter} algorithm scans the \mjj\ region above 1.3~\TeV\ as indicated by the vertical dashed line. The vertical solid lines refer to the boundaries of the most significant interval flagged by \texttt{BumpHunter}; this is found to be $[1.92,2.11]$~\TeV\ for which the $p$-value is stated in the figure. Open squares (circles) are the signal distributions of the 1.6 (2.5)~\TeV\ LUV \Zprime\ boson on top of the data with the cross sections inflated by a factor of 20 (50). The lower panel shows the significance calculated using the background estimate and data. The significance is defined as $(d_{i} - b_{i})/\delta$, where $d_{i}$ refers to the data count while $b_{i}$ refers to the background estimate, and $\delta$ is an uncertainty taking both the uncertainty on the data count and background estimate into account~\cite{Choudalakis_2012}.}
\label{fig:fit_plot}
\end{figure}
 
A binned likelihood is constructed with the signal templates obtained from
simulation and the background template given by FD. The likelihood function,
$\mathcal{L}(\mu,\vec{\theta})$, is a product of Poisson probability terms over
all the \mjj bins, where the parameter of interest, $\mu$, is defined as the
cross section, $ \sigma(pp\to b\bar{b}Z^{\prime})$, times the branching ratio,
${B}(Z^{\prime}\to b\bar{b}) $, of the LUV \Zprime boson, and a set of nuisance
parameters (NPs), $\vec{\theta}$, is used to encode systematic uncertainties in
the signal and background expectations. The background normalization and shape
are estimated by FD prior to the likelihood fit, and corresponding
uncertainties are included as NPs in the likelihood function.  The NPs allow
variations of the expectations for signal and background according to the
systematic uncertainties, subject to Gaussian constraints in the likelihood
fit.  The maximum-likelihood fit to the observed \mjj spectrum is performed
independently for each LUV \Zprime boson mass hypothesis with the starting
point of the \mjj spectrum set to 1~\TeV\ so that the majority of signal events
are included.
 
Since the observed data are consistent with the smoothly falling background
expectation, upper limits on the production cross section of the LUV \Zprime\
model are obtained at 95\% confidence level (CL) using the
CL$_{\textrm{s}}$~\cite{Cowan:2010js,Read:2002,Junk:1999kv} method. Various
signal mass hypothesis tests and CL$_{\textrm{s}}$ values are calculated using a
frequentist framework, HistFitter~\cite{HistFitter:2014}, which is an interface
to the RooFit package~\cite{Verkerke2003TheRT}, with a profile likelihood
ratio, $q_{\mu} = -2{\rm
ln}(\mathcal{L}(\mu,\vec{\theta}_{\mu})/\mathcal{L}(\hat{\mu},\hat{\vec{\theta}}_{\hat{\mu}}))$,
as the test statistic. The values of $\hat{\mu}$ and
$\hat{\vec{\theta}}_{\hat{\mu}}$ maximize the likelihood function, while
$\vec{\theta}_{\mu}$ maximizes the likelihood function for a given value of
$\mu$. For each signal mass hypothesis, a CL$_{\textrm{s}}$ value is evaluated using
the asymptotic formula~\cite{Cowan:2010js}. The predicted LUV \Zprime boson
production cross sections times the branching ratio to $b\overline{b}$ are
overlaid in Figure~\ref{fig:limit}. At 3~\TeV, where the number of events is
small, the results from pseudoexperiments are consistent with the ones using
the asymptotic formula. The difference in the mean is at the few-percent level
and that in the uncertainty band is about 10\%. A coverage test using an
ensemble of pseudodata samples is performed to validate the overall
implementation of systematic uncertainties. When signal events are injected
with a strength corresponding to the mean cross section excluded at 95\% CL
obtained by considering the background-only pseudodata samples, the probability
of excluding the given signal is less than 5\% for a range of signal masses spanning
1.3~\TeV\ to 3~\TeV. An additional cross-check is performed to validate the
robustness of the complete analysis procedure against possible signal
contamination. This cross-check, which considers two LUV \Zprime\ masses
(1.3~\TeV\ and 3~\TeV), is carried out by subtracting signal events
corresponding to the expected limit from data, and creating a new set of
pseudodata samples using the signal-subtracted data. The allowed range of $N$
for FD and the uncertainties associated with FD, such as the fit parameter
uncertainty, spurious signal uncertainty and the fit bias uncertainty, are
rederived and then applied in the statistical analysis of the original data.
The resulting limits on $N$ are the same as the ones obtained in
Section~\ref{sec:bkg} and the expected and observed limits are shifted by less
than 2\% compared to the nominal result.
 
\begin{figure}[!ht]
\centering
\includegraphics[scale=0.5]{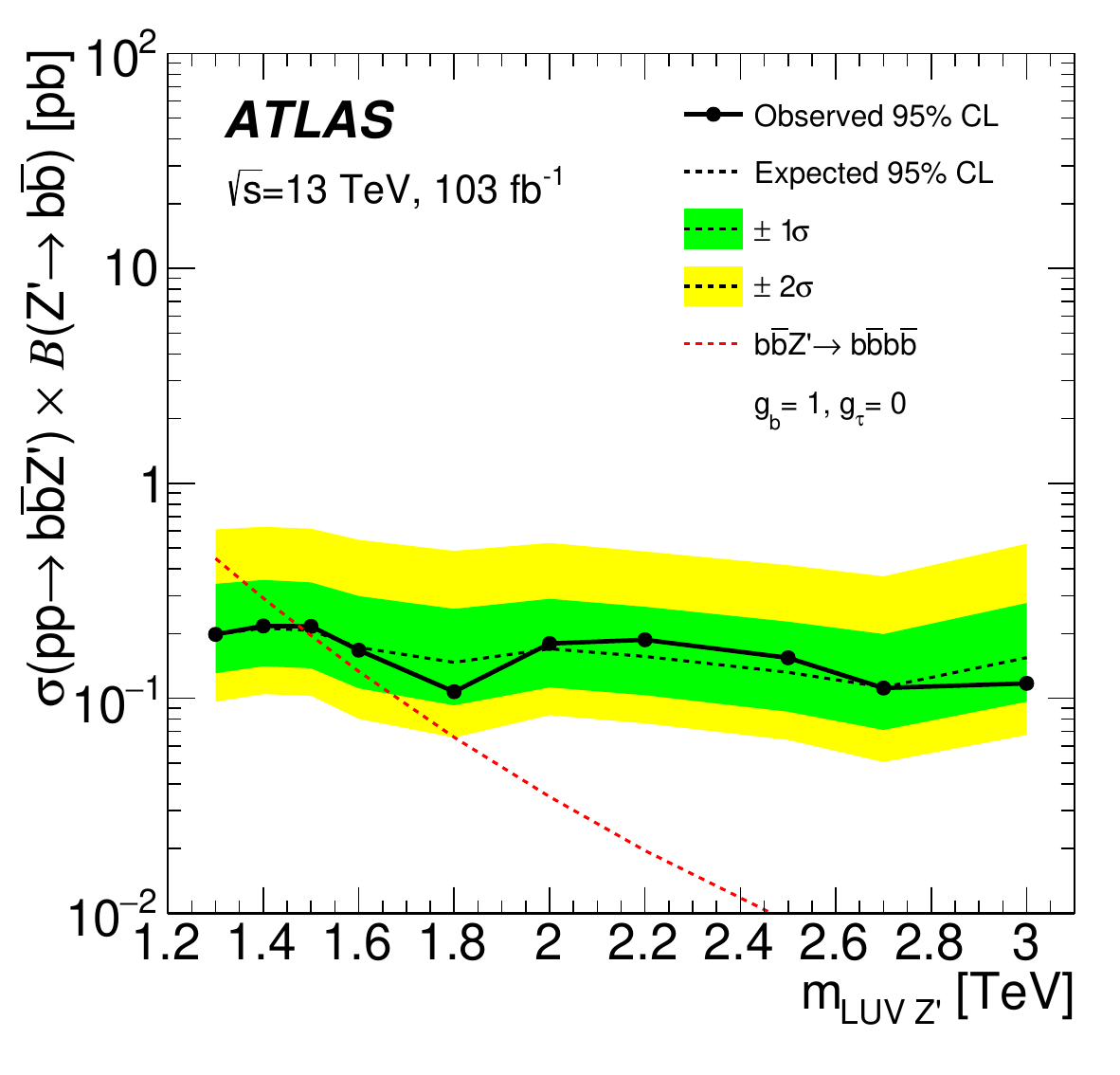}
\caption{The observed (solid) and expected (dashed) 95\% CL upper limits on
the production of $Z^{\prime}\to b\bar{b}$ in association with $b$-quarks. The bands surrounding the expected limit show the 68\% and 95\% confidence intervals. The predicted cross sections of the LUV \Zprime model times the  branching ratio to $b\overline{b}$, with $g_{b}$ set to unity and $g_{\tau}$ set to zero, calculated at LO with \MGMCatNLO, are also overlaid.}
\label{fig:limit}
\end{figure}

% End of text imported from the .//chapters/Interpretation.tex input file

% The next lines are included from the .//chapters/Conclusion.tex input file
\FloatBarrier
\section{Conclusion}
\label{sec:conclusion}
 
A search for new heavy particles in the final state with the two leading jets
$b$-tagged and at least one additional $b$-tagged jet among the third and
fourth jets is performed in a previously uncovered \mjj\ region using 103 ~\ifb\
of $pp$ collision data recorded at $\sqrt{s} = 13$~\TeV\ with the ATLAS
detector at the LHC during 2017 and 2018. The \mjj\ spectrum of the
two leading $b$-tagged jets is analyzed with an innovative background
estimation method, functional decomposition, utilizing a truncated series. The
observed data are compatible with the hypothesis of a smoothly falling
background within the \mjj\ region from 1.3~\TeV\ to 3.6~\TeV. Upper limits on
the production of lepton-universality-violating \Zprime\ bosons times the
branching ratio to $b\overline{b}$ are derived, and exclude such \Zprime\ bosons
with masses between 1.3 and 1.45~\TeV\ at 95\% CL.

% End of text imported from the .//chapters/Conclusion.tex input file

\section*{Acknowledgments}

% The next lines are included from the .//acknowledgements/Acknowledgements.tex input file

We thank CERN for the very successful operation of the LHC, as well as the
support staff from our institutions without whom ATLAS could not be
operated efficiently.
 
We acknowledge the support of
ANPCyT, Argentina;
YerPhI, Armenia;
ARC, Australia;
BMWFW and FWF, Austria;
ANAS, Azerbaijan;
SSTC, Belarus;
CNPq and FAPESP, Brazil;
NSERC, NRC and CFI, Canada;
CERN;
ANID, Chile;
CAS, MOST and NSFC, China;
Minciencias, Colombia;
MSMT CR, MPO CR and VSC CR, Czech Republic;
DNRF and DNSRC, Denmark;
IN2P3-CNRS and CEA-DRF/IRFU, France;
SRNSFG, Georgia;
BMBF, HGF and MPG, Germany;
GSRI, Greece;
RGC and Hong Kong SAR, China;
ISF and Benoziyo Center, Israel;
INFN, Italy;
MEXT and JSPS, Japan;
CNRST, Morocco;
NWO, Netherlands;
RCN, Norway;
MNiSW and NCN, Poland;
FCT, Portugal;
MNE/IFA, Romania;
JINR;
MES of Russia and NRC KI, Russian Federation;
MESTD, Serbia;
MSSR, Slovakia;
ARRS and MIZ\v{S}, Slovenia;
DSI/NRF, South Africa;
MICINN, Spain;
SRC and Wallenberg Foundation, Sweden;
SERI, SNSF and Cantons of Bern and Geneva, Switzerland;
MOST, Taiwan;
TAEK, Turkey;
STFC, United Kingdom;
DOE and NSF, United States of America.
In addition, individual groups and members have received support from
BCKDF, CANARIE, Compute Canada and CRC, Canada;
COST, ERC, ERDF, Horizon 2020 and Marie Sk{\l}odowska-Curie Actions, European Union;
Investissements d'Avenir Labex, Investissements d'Avenir Idex and ANR, France;
DFG and AvH Foundation, Germany;
Herakleitos, Thales and Aristeia programmes co-financed by EU-ESF and the Greek NSRF, Greece;
BSF-NSF and GIF, Israel;
Norwegian Financial Mechanism 2014-2021, Norway;
La Caixa Banking Foundation, CERCA Programme Generalitat de Catalunya and PROMETEO and GenT Programmes Generalitat Valenciana, Spain;
G\"{o}ran Gustafssons Stiftelse, Sweden;
The Royal Society and Leverhulme Trust, United Kingdom.
 
The crucial computing support from all WLCG partners is acknowledged gratefully, in particular from CERN, the ATLAS Tier-1 facilities at TRIUMF (Canada), NDGF (Denmark, Norway, Sweden), CC-IN2P3 (France), KIT/GridKA (Germany), INFN-CNAF (Italy), NL-T1 (Netherlands), PIC (Spain), ASGC (Taiwan), RAL (UK) and BNL (USA), the Tier-2 facilities worldwide and large non-WLCG resource providers. Major contributors of computing resources are listed in Ref.~\cite{ATL-SOFT-PUB-2021-003}.

% End of text imported from the .//acknowledgements/Acknowledgements.tex input file

\clearpage

\printbibliography

\clearpage \input{atlas_authlist}

\end{document}

%% file: atlas_authlist.tex
% ATLAS Collaboration author list
% Reference date of EXOT-2018-09 is 2021-03-10
% Author list last updated on date 26-OCT-21
% Data extracted on 26-Oct-2021 for paper reference EXOT-2018-09
% at 11:50am
 
\begin{flushleft}
\hypersetup{urlcolor=black}
{\Large The ATLAS Collaboration}

\bigskip

\AtlasOrcid[0000-0002-6665-4934]{G.~Aad}$^\textrm{\scriptsize 99}$,    
\AtlasOrcid[0000-0002-5888-2734]{B.~Abbott}$^\textrm{\scriptsize 125}$,    
\AtlasOrcid[0000-0002-7248-3203]{D.C.~Abbott}$^\textrm{\scriptsize 100}$,    
\AtlasOrcid[0000-0002-2788-3822]{A.~Abed~Abud}$^\textrm{\scriptsize 34}$,    
\AtlasOrcid[0000-0002-1002-1652]{K.~Abeling}$^\textrm{\scriptsize 51}$,    
\AtlasOrcid[0000-0002-2987-4006]{D.K.~Abhayasinghe}$^\textrm{\scriptsize 91}$,    
\AtlasOrcid[0000-0002-8496-9294]{S.H.~Abidi}$^\textrm{\scriptsize 27}$,    
\AtlasOrcid[0000-0001-5329-6640]{H.~Abramowicz}$^\textrm{\scriptsize 158}$,    
\AtlasOrcid[0000-0002-1599-2896]{H.~Abreu}$^\textrm{\scriptsize 157}$,    
\AtlasOrcid[0000-0003-0403-3697]{Y.~Abulaiti}$^\textrm{\scriptsize 5}$
% ...
\bigskip
\\

% ...
$^{5}$High Energy Physics Division, Argonne National Laboratory, Argonne IL; United States of America.\\
% ...
$^{27}$Physics Department, Brookhaven National Laboratory, Upton NY; United States of America.\\
% ...
$^{34}$CERN, Geneva; Switzerland.\\
% ...
$^{51}$II. Physikalisches Institut, Georg-August-Universit\"{a}t G\"ottingen, G\"ottingen; Germany.\\
% ...
$^{91}$Department of Physics, Royal Holloway University of London, Egham; United Kingdom.\\
% ...
$^{99}$CPPM, Aix-Marseille Universit\'e, CNRS/IN2P3, Marseille; France.\\
$^{100}$Department of Physics, University of Massachusetts, Amherst MA; United States of America.\\
% ...
$^{125}$Homer L. Dodge Department of Physics and Astronomy, University of Oklahoma, Norman OK; United States of America.\\
% ...
$^{157}$Department of Physics, Technion, Israel Institute of Technology, Haifa; Israel.\\
$^{158}$Raymond and Beverly Sackler School of Physics and Astronomy, Tel Aviv University, Tel Aviv; Israel.\\
% ...

\end{flushleft}

% Created with Glance <Atlas.Glance@cern.ch>